\pdfoutput=1
\documentclass[11pt,a4paper]{article}

\usepackage[T1]{fontenc}
\usepackage[utf8]{inputenc}
\usepackage{amsmath}
\usepackage[tt=false]{libertine}
\usepackage[libertine]{newtxmath}
\usepackage[varqu]{zi4}
\usepackage[margin=2.5cm]{geometry}
\usepackage{microtype}
\usepackage{graphicx}
\usepackage{xcolor}
\usepackage{booktabs}
\usepackage[numbers,sort&compress]{natbib}
\usepackage{authblk}

\usepackage[font=small,labelfont=bf]{caption}
\usepackage{xurl}
\usepackage[colorlinks=true,linkcolor=blue!50!black,citecolor=blue!50!black,urlcolor=blue!50!black]{hyperref}
\usepackage[section]{placeins}

\usepackage{fancyhdr}
\usepackage{tikz}
\usetikzlibrary{arrows.meta,positioning,calc,patterns,decorations.pathreplacing}
\usepackage{array}
\newcolumntype{L}[1]{>{\raggedright\arraybackslash}p{#1}}
\newcolumntype{Y}{>{\centering\arraybackslash}p{0.95cm}}

\DeclareRobustCommand{\fullc}{\tikz[baseline=-0.6ex]\fill (0,0) circle (2.3pt);}
\DeclareRobustCommand{\halfc}{\tikz[baseline=-0.6ex]{\draw (0,0) circle (2.3pt); \fill (0,0) -- (90:2.3pt) arc (90:270:2.3pt) -- cycle;}}
\DeclareRobustCommand{\nonec}{\tikz[baseline=-0.6ex]\draw (0,0) circle (2.3pt);}

\newif\ifshownotes
\shownotesfalse

\title{Towards Elastic Least Privilege for Agent Planning}
\author[1]{Simon Parkinson\thanks{Corresponding author: \href{mailto:s.parkinson@mmu.ac.uk}{s.parkinson@mmu.ac.uk}}}
\author[2]{Zhipeng Wang}
\author[2]{Mustafa Mustafa}
\affil[1]{School of Computing and Mathematics, Manchester Metropolitan University, Manchester, United Kingdom}
\affil[2]{Systems and Software Security, University of Manchester, Manchester, United Kingdom}
\date{}

\begin{document}

\maketitle
\thispagestyle{fancy}

\begin{abstract}
General-purpose AI agents require authority that changes between tasks and during execution. Common deployments trade broad permissions that expose unnecessary authority and narrow permissions that raise approval requests. Existing mechanisms can enforce dynamic authority, but agent tool interfaces rarely provide the agent's planner, the component that chooses its actions, a means of comparing the authority required by alternative actions or of seeing the cost of the authority it already holds. Drawing on the rich history of access control research, we propose the concept of \emph{elastic least privilege}, in which an agent begins with a small baseline of authority and may acquire more within an owner-defined envelope. Elevations are leased and measured against a replenishing budget. Their cost increases with power, proximity to mandatory safety invariants, and repeated requests. Costs are visible to the planner, which can prefer lower-privilege choices and release authority it no longer needs. Authority returns automatically to the baseline when a lease expires, when it is released, or when the budget is exhausted, while persistent legitimate demand may prompt an owner-approved change to the baseline. Building on prior risk-priced and budget-aware access control, we redesign these mechanisms for agents. How best to implement the approach and how well it works in practice are open research questions. In this paper, we present how the approach could work, illustrate the model, and set an open research agenda.
\end{abstract}

\noindent\textbf{Keywords:} least privilege, AI agents, over-privilege, risk-based access control, privilege budgets, just-in-time access, Model Context Protocol

\section{Introduction}
\label{sec:intro}

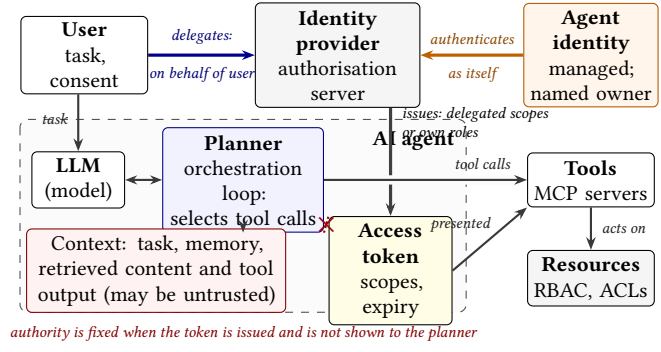
\begin{figure}[tbp]
\centering
\begin{tikzpicture}[font=\scriptsize,
  box/.style={draw, rounded corners=2pt, align=center, fill=white, inner sep=2.5pt},
  arr/.style={-{Stealth[length=1.7mm]}, thick, gray!45!black},
  dlg/.style={-{Stealth[length=1.7mm]}, very thick, blue!55!black},
  own/.style={-{Stealth[length=1.7mm]}, very thick, orange!75!black},
  lab/.style={font=\tiny\itshape, text=gray!25!black, align=center}]
  \node[box, text width=1.5cm] (user) at (0.8,0) {\textbf{User}\\ task, consent};
  \node[box, text width=2.0cm, fill=gray!8] (idp) at (4.15,0) {\textbf{Identity provider}\\ authorisation server};
  \node[box, draw=orange!70!black, fill=orange!8, text width=1.6cm] (aid) at (7.5,0) {\textbf{Agent identity}\\ managed; named owner};
  \draw[rounded corners=3pt, dashed, draw=gray!60!black, fill=gray!3] (-0.05,-0.85) rectangle (5.85,-3.4);
  \node[anchor=north east, font=\scriptsize\bfseries] at (5.83,-0.87) {AI agent};
  \node[box, text width=1.05cm] (llm) at (0.72,-1.65) {\textbf{LLM}\\ (model)};
  \node[box, draw=blue!50!black, fill=blue!5, text width=1.95cm] (plan) at (2.9,-1.65) {\textbf{Planner}\\ orchestration loop:\\ selects tool calls};
  \node[box, draw=red!50!black, fill=red!5, text width=3.25cm] (ctx) at (1.75,-2.85) {Context: task, memory, retrieved content and tool output (may be untrusted)};
  \node[box, fill=yellow!12, text width=1.45cm] (tok) at (4.85,-2.85) {\textbf{Access token}\\ scopes, expiry};
  \node[box, text width=1.5cm] (tools) at (7.5,-1.65) {\textbf{Tools}\\ MCP servers};
  \node[box, fill=gray!8, text width=1.5cm] (res) at (7.5,-2.95) {\textbf{Resources}\\ RBAC, ACLs};
  \draw[dlg] (user.east) -- (idp.west) node[lab, text=blue!45!black, midway, above] {delegates:}  node[lab, text=blue!45!black, midway, below] {on behalf of user};
  \draw[own] (aid.west) -- (idp.east) node[lab, text=orange!55!black, midway, above] {authenticates}  node[lab, text=orange!55!black, midway, below] {as itself};
  \draw[arr, very thick, black!75] (idp.south -| tok.north) -- (tok.north) node[lab, text=black, pos=0.13, right, align=left] {issues: delegated scopes\\ or own roles};
  \draw[arr] (user.south -| llm.north) -- (llm.north) node[lab, pos=0.42, left] {task};
  \draw[thick, gray!45!black, {Stealth[length=1.6mm]}-{Stealth[length=1.6mm]}] (llm) -- (plan);
  \draw[arr] (ctx.north -| plan.south) -- (plan.south);
  \draw[arr, preaction={draw, gray!3, line width=3.5pt}] (plan.east) -- (tools.west) node[lab, pos=0.78, above] {tool calls};
  \draw[arr] (tok.east) -- (tools.south west) node[lab, pos=0.55, above left, inner sep=1pt] {presented};
  \draw[arr] (tools) -- (res) node[lab, midway, right] {acts on};
  \draw[red!60!black, densely dotted, thick] (plan.south east) -- (tok.north west) node[midway, font=\bfseries, text=red!60!black] {$\times$};
  \node[lab, text=red!50!black, anchor=north] at (2.9,-3.45) {authority is fixed when the token is issued and is not shown to the planner};
\end{tikzpicture}
\vspace{-0.8\baselineskip}
\caption{Conceptual view of an AI agent and permissioning}
\label{fig:concept}
\end{figure}

Artificial intelligence (AI) agents combine large language models (LLMs) with tools, credentials, and an objective, then choose which actions to take in pursuit of that objective. The Model Context Protocol (MCP) standardises how agents discover and invoke tools, and a recent study now counts MCP servers in the tens of thousands~\cite{chen2026rethinkingmcp}. Agents operate on top of existing deployments, reaching resources through access controls that were designed and implemented for people and services. A tool exposes operations whose exercise depends on the authority available to the agent. An agent may have its own identity or may act on behalf of another principal. Figure~\ref{fig:concept} illustrates the concept, where the planner turns the user's task into tool calls, and the tools act on resources that enforce their own access control. Authority reaches the agent, either by delegation from the user so that the agent acts on their behalf (blue), or issued to the agent's own managed identity (orange). In both cases, it is fixed when issued and is not visible to the planner.

The principle of {\it least privilege} requires that resources are accessed only with the privileges necessary to complete the job~\cite{saltzer1975protection}, but in practice it is rarely achieved. For example, identities in one vendor's multi-cloud analysis used only 1\% of their permissions~\cite{chik2023entra}, 99.8\% of 447,000 GitHub workflows were over-privileged~\cite{koishybayev2022github}, and excess authority persists because most of it is never exercised by those who hold it~\cite{parkinson2017ransomware}. Although the principle is well established in access control, it is especially important for agents, which, like users, obtain permissions for multiple resources through many different control mechanisms. Industry best-practice guidance recommends that each agent be a {\it first-class principal}, with a dedicated agent identity and a specific scope~\cite{microsoft2026leastprivilege}. However, this is often hard to realise because general-purpose agents serve heterogeneous tasks that combine natural-language tasking, runtime tool selection, cross-system composition, and exposure to untrusted content. Their required authority is task-dependent and may change as execution reveals new subgoals, so a single allocated permission set sufficient for all supported tasks will often exceed what any particular task requires. 

Common deployment patterns therefore trade over-entitlement against runtime friction. A broad grant supports task completion, but leaves unnecessary authority available to an injected instruction. For example, a malicious instruction in a public repository has been reported to steer an agent to read its user's private repositories and publish them~\cite{milanta2025githubmcp}. This combination of private data, untrusted content and external communication, the ``lethal trifecta''~\cite{willison2025trifecta}, is a property of what is held together, not of any single call. A narrow allocation with approval for each extension reduces authority, but transfers the burden to users, who are likely to approve broad increases in authority in order to make progress. One vendor reports that users of its coding agent approved 93\% of permission prompts~\cite{hughes2026automode}. Approval decisions are also rarely graded by what the agent already holds or how often it has asked, and LLM monitors that might replace them can themselves be subverted~\cite{terekhov2026adaptive}. Task-scoped credentials, policy-based automatic elevation, pre-authorised envelopes and per-task enforcement outside the model~\cite{shi2025progent,debenedetti2025camel,costa2025fides,palumbo2026formal,li2026ace} mitigate this trade-off. However, the authority estimated for a task can be too narrow or too broad~\cite{meng2026cellmate}, and, to the best of our knowledge, none of these mechanisms turns privilege into a signal the \emph{planner} can use, by which we mean the component of an agent, typically an LLM within an orchestration loop, that decomposes a task and selects which tool calls to make and in what order. Policy systems may expose scopes, decision reasons and risk labels, but these describe a single request when it is made. What is missing is a planning signal with three parts, namely a state-dependent marginal cost that compares the additional authority each alternative would require given what is already held, an ongoing holding cost for authority that is held but unused, and an explicit release action whose effect is guaranteed by enforcement that does not depend on the planner. Without such a signal, privilege is enforced at execution time rather than represented as a variable during planning, and agents often choose an over-privileged tool when a lower-privilege one would suffice. Across 544 scenarios, 6 of 11 LLM agents did so in more than 30\% of cases~\cite{yang2026toolprivbench}. 

Consider, as a running example, a coding assistant connected to GitHub through the GitHub MCP server. If the user authenticates with a classic personal access token carrying the \texttt{repo} scope, the agent can use about 40 tools, 16 of which can write~\cite{github2026mcpserver}, in every repository the user can access~\cite{github2026pat,github2026scopes}. Asked to summarise the open issues of one repository, it needs two read-only tools on that repository. However, the token also gives it all other tools in every other repository, which is the authority that the injected instruction described above exploited~\cite{milanta2025githubmcp}. A fine-grained token restricted to read access on that repository would have prevented the attack, but the user's next task, fixing a failing build, needs to write code, and the one after may need another repository. The example captures the central dilemma that useful agents need authority that can change, but more consequential changes must encounter greater resistance, and the planner would benefit from knowing when it is asking for or holding more than it needs.

\begin{figure}[tbp]
\centering
\begin{tikzpicture}[font=\scriptsize, x=8cm, y=1cm]
  \fill[green!8] (0,0) rectangle (0.18,3.3);
  \fill[blue!6] (0.18,0) rectangle (0.62,3.3);
  \fill[orange!10] (0.62,0) rectangle (0.9,3.3);
  \fill[pattern=north east lines, pattern color=red!45] (0.9,0) rectangle (1.0,3.3);
  \draw[red!60!black, line width=1pt] (0.9,0) -- (0.9,3.3);
  \draw[blue!50!black, dashed] (0.62,0) -- (0.62,3.3);
  \draw[-{Stealth[length=2mm]}] (0,0) -- (1.03,0);
  \node[anchor=north east, align=right] at (1.03,-0.02) {authority held (power, proximity)};
  \draw[-{Stealth[length=2mm]}] (0,0) -- (0,3.45) node[anchor=south west, xshift=-2pt] {resistance (cost rate)};
  \begin{scope}
  \clip (0,0) rectangle (0.9,3.3);
  \draw[thick, blue!50!black, domain=0.18:0.62, samples=40] plot (\x, {0.6*(\x-0.18)/(0.9-\x)});
  \draw[thick, blue!50!black, dotted, domain=0.62:0.89, samples=60] plot (\x, {0.6*(\x-0.18)/(0.9-\x)});
  \draw[thick, orange!70!black, dashed, domain=0.18:0.62, samples=40] plot (\x, {1.2*(\x-0.18)/(0.9-\x)});
  \end{scope}
  \draw[very thick, green!35!black] (0,0) -- (0.18,0);
  \node[align=center, text=green!25!black] at (0.09,2.8) {\textbf{rest}\\ baseline\\ $B(a)$,\\ no cost};
  \node[align=center, text=blue!40!black] at (0.40,2.8) {\textbf{elastic}\\ within envelope $V(a)$:\\ costed, leased, recoils};
  \node[align=center, text=orange!45!black] at (0.765,1.0) {\textbf{rigid}\\ only the owner\\ moves $B$ or $V$};
  \node[align=center, text=red!45!black, fill=white, inner sep=1pt] at (0.95,1.5) {\textbf{fracture}\\ ceiling\\ $C$};
  \draw[thick, blue!50!black] (0.21,1.95) -- (0.27,1.95) node[anchor=west, text=black] {first request};
  \draw[thick, orange!70!black, dashed] (0.21,1.65) -- (0.27,1.65) node[anchor=west, text=black] {repeated requests};
  \draw[-{Stealth[length=1.6mm]}, thick, gray!40!black] (0.12,-0.75) -- (0.47,-0.75) node[midway, below] {stretch (request, paid from budget)};
  \draw[-{Stealth[length=1.6mm]}, thick, gray!40!black] (0.47,-1.3) -- (0.12,-1.3) node[midway, below] {recoil (lease expiry, release or exhaustion)};
\end{tikzpicture}
\vspace{-0.8\baselineskip}
\caption{Illustration of the elastic least privilege concept}
\label{fig:elastic}
\end{figure}
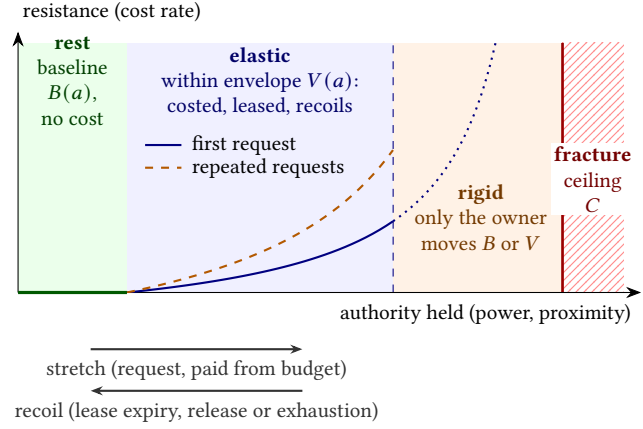

Least privilege can be enforced without the planner's awareness, but a planner that chooses among alternative action sequences cannot reliably optimise for privilege unless the authority implications of its alternatives are represented in a form it can observe. In this BlueSky paper, we propose \emph{elastic least privilege} for AI agents (Figure~\ref{fig:elastic}). %
The approach addresses the trade-off between narrow and broad authority and the tendency of users to grant broad permissions. An agent holds a small \emph{baseline} of authority at rest. When a task needs more, the agent may request additional authority within an \emph{envelope} set in advance by its owner, but each extension incurs a \emph{cost} that grows with the power of the authority held and becomes infinite at a \emph{ceiling} of combinations that must never be held. Elevated authority \emph{recoils}, that is, returns to the baseline, when its lease expires. Repeated requests increase the cost of further requests, and persistent demand prompts the owner to consider changing the baseline. Costs are drawn from a \emph{budget} that replenishes over time and are shown to the planner, which sees what each action would cost and what unused authority is costing it, and can release that authority.

The concept is based on a substantial body of prior work leveraged from the access control community. Risk-priced access was proposed for the intelligence community in 2004~\cite{jason2004horizontal}, and further developments of similar approaches were undertaken through risk tokens, markets, and budget-aware role-based access control (RBAC)~\cite{cheng2007fuzzymls,molloy2008trading,liu2010budget,salim2013brbac}. These schemes bounded the priced access a principal could consume and used prices to steer people towards less powerful options~\cite{salim2013brbac}. A planner can consider a cost on every action without suffering decision fatigue, but has no inherent reason to conserve its budget, so steering must be supported through an appropriate optimisation criterion. Our proposition is that authority held over time can be exposed to the planner as an action-dependent cost, while an independent issuer enforces history-sensitive ceilings, cumulative exposure bounds, and automatic recoil. 

The remainder of the paper is organised as follows. Section~\ref{sec:ancestry} reviews prior work on priced and budgeted access control, Section~\ref{sec:model} presents the model, and Section~\ref{sec:example} a worked example. Section~\ref{sec:limits} discusses limitations and Section~\ref{sec:agenda} sets out an open research agenda.

\section{Prior and Related Work}
\label{sec:ancestry}

Associating a cost with access according to its risk is not a new concept. It was proposed for human users more than two decades ago, and this section reviews that work and the work on agent security closest to this article (Table~\ref{tab:ancestry}), and then explains how agents alter its assumptions.

\begin{table*}[tbp]
\caption{Prior approaches and their coverage of the features of the elastic model. Abbreviations are multi-level security (MLS), budget-aware role-based access control (B-RBAC), just-in-time (JIT) and separation of duty (SoD).}
\label{tab:ancestry}
\footnotesize
\setlength{\tabcolsep}{2.6pt}
\resizebox{\textwidth}{!}{%
\begin{tabular}{@{}L{2.75cm}L{2.65cm}YYL{1.55cm}YYYL{1.95cm}YY@{}}
\toprule
Approach & What is charged & Ceiling & Cost depends on what is held & Budget refill & Cost visible to requester & Leases, recoil & History raises cost & When the budget runs out & Moves the baseline & Signals over-privilege \\
\midrule
JASON 2004~\cite{jason2004horizontal} (people) & Each access, by risk, form and duration & \fullc & \nonec & Annual (suggested) & \nonec & \halfc & \halfc & Ask manager for more & \halfc & \nonec \\
Fuzzy MLS, risk markets 2007--08~\cite{cheng2007fuzzymls,molloy2008trading} (people) & Risk above a soft boundary & \fullc & \nonec & Per interval, or bought & \fullc & \nonec & \nonec & Buy tokens & \nonec & \nonec \\
Budget-based 2010~\cite{liu2010budget} (people) & Exceptional access & \nonec & \nonec & Periodic reset & \halfc & \nonec & \nonec & User punished & \nonec & \nonec \\
B-RBAC 2012--13~\cite{salim2013brbac,salim2012thesis} (people) & Each execution and escalation & \fullc & \halfc & Each round & \halfc & \halfc & \halfc & Flag for audit & \halfc & \halfc \\
JIT elevation~\cite{microsoft2026pim,apono2026guard} (people, agents) & Nothing; approval by role or tier & \halfc & \nonec & -- & \nonec & \fullc & \nonec & -- & \nonec & \nonec \\
Progent 2025~\cite{shi2025progent} (agents) & Nothing; expansion needs approval & \fullc & \nonec & -- & \nonec & \nonec & \nonec & -- & \nonec & \nonec \\
Irreversibility Budget 2026~\cite{mohammadi2026irreversibility} (agents) & Each effect, by residual loss & \nonec & \nonec & Windowed & \nonec & \nonec & \nonec & Exception allocation & \nonec & \nonec \\
\textbf{Elastic (this paper)} (agents) & Elevated authority held, per minute & \fullc & \fullc & Continuous & \fullc & \fullc & \fullc & One batched owner request & \fullc & \fullc \\
\bottomrule
\end{tabular}}
\end{table*}

The JASON report proposed that risk be ``tokenised'', connecting one token to the ``risk associated with one-day, soft-copy-only access to one document by the average Secret-cleared individual''~\cite{jason2004horizontal}. Riskier transactions cost more tokens, which encourages the use of less risky access. These tokens can be refreshed annually, and an employee whose risk factor increases requires management approval~\cite{jason2004horizontal}. Fuzzy multi-level security (MLS) introduced a risk-adaptive gap between allow and deny~\cite{cheng2007fuzzymls}, and a market in risk tokens allows employees to evaluate a cost before purchasing access~\cite{molloy2008trading}. Budget-based access control assigned periodic risk budgets and charged for exceptional access by its risk classification and the requester's risk score~\cite{liu2010budget}. Budget-aware RBAC (B-RBAC) is the most complete design~\cite{salim2013brbac,salim2012thesis}, in which users pay for permissions from a budget that expires each round. Escalation carries a cost multiplier, which is infinite where separation of duty forbids it, and role weight is charged so that users prefer cheaper roles, which ``indirectly connects the observance of the principle of least privilege with users' utility''~\cite{salim2013brbac}. Risk- and trust-aware RBAC~\cite{dimmock2004trustrisk,chen2012riskrbac,baracaldo2012trustrisk,bijon2012risksessions} and MarketNet~\cite{yemini2000marketnet} are related and are based on similar conceptual approaches.

For agents, privileged access management solutions grant a time-limited elevation, automatically, with approval or not at all, by role or sensitivity level~\cite{microsoft2026pim,apono2026guard}. Notable examples include Progent, which narrows an agent's policy automatically and expands it only with approval~\cite{shi2025progent}. Risk-aware causal gating penalises high-risk tools in path selection~\cite{iyer2026racg}, and the Irreversibility Budget measures each effect against windowed budgets~\cite{mohammadi2026irreversibility}. Together, these works established most of the economic and enforcement bases of our proposal. More specifically, they established that risk can be denominated in a common unit, that a budget bounds the amount of priced access a principal can consume in a period, whatever its intent, that prices can steer principals towards less powerful options, that exhaustion is a useful signal for audit, and that enforcement belongs below the agent. 

As Table~\ref{tab:ancestry} shows, what none of these approaches provides is a way to tell the planner what each choice would cost in privilege and what the over-privilege it already holds is costing it, with the option to release it. The concept presented in this paper builds on these ideas, and the bound we state in Section~\ref{sec:guarantees} is the budget bound of B-RBAC restated for the authority held over time.  

Many have identified obstacles and shortcomings in previous work on pricing access control~\cite{jason2004horizontal,salim2013brbac,salim2012thesis}. Although these concerns are significant, we argue that agentic systems change the environment and provide benefits that outweigh prior concerns. Decision overhead was the first concern, as B-RBAC's authors raised concerns about the burden that budget decisions would impose on users~\cite{salim2012thesis}, and JASON recommended that tokens be supplied to people with minimal disruption~\cite{jason2004horizontal}. Unlike a human user, a planner can be presented with a structured cost for every candidate request without generating repeated approval interactions, although whether it uses those costs reliably remains an empirical question. Budget periods and costing raised a second concern, as B-RBAC's authors asked whether users would spend all of their budget before a budget round ended to avoid losing it. They also found assigning a cost to every task a barrier~\cite{salim2013brbac}. An agent's budget can operate on fine-grained temporal windows, and costs can attach to operation classes that platforms already use (Section~\ref{sec:stiffness}). Incentive was the third concern, since the steering of people depended on self-interest. Users were expected to preserve their budget on the grounds that the faster the budget is exhausted, the higher the likelihood that they will be audited~\cite{salim2013brbac}. Delegating these decisions to an agent removes such tactical spending by people. However, an agent has no fear of audit, so its budget must be enforced rather than relied upon as an incentive.

\section{Elastic Least Privilege}
\label{sec:model}

Next, we specify the proposal sufficiently to define its threat model and state its guarantees. In terms of terminology, a \emph{permission} is an authorised operation on a resource, represented here as an operation--resource pair. A \emph{grant} is a policy decision that confers one or more permissions. A credential represents or enables the exercise of granted permissions. The \emph{effective authority} of an agent at time $t$ is the set of operation--resource pairs that it can cause the deployment to perform at $t$, directly or through a credential or tool server in its reach. The issuer conservatively accounts for this authority using the permissions represented by active grants. We use \emph{elevated authority} for effective authority above the baseline. For costing, permissions are grouped into \emph{authority units} $x$, each an operation class on a resource, such as writing the contents of one repository, and each elevated grant $g$ confers one unit $x_g$. The full notation used in this work is given in Table~\ref{tab:notation}.

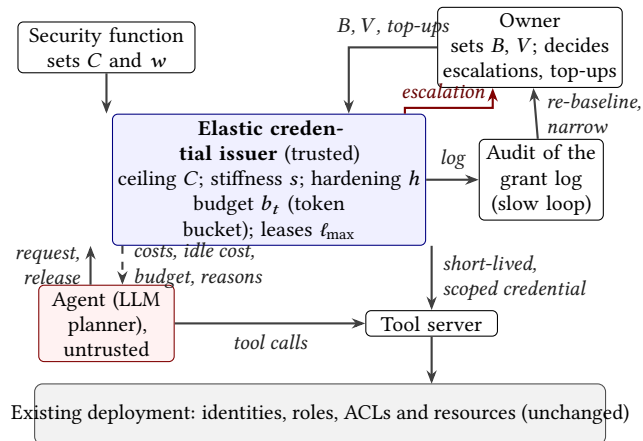
\begin{figure}[tbp]
\centering
\begin{tikzpicture}[font=\scriptsize,
  box/.style={draw, rounded corners=2pt, align=center, fill=white, inner sep=2pt},
  arr/.style={-{Stealth[length=1.8mm]}, thick, gray!50!black},
  lab/.style={font=\scriptsize\itshape, text=gray!25!black}]
  \node[box, text width=2.2cm] (sec) at (0,0) {Security function\\ sets $C$ and $w$};
  \node[box, text width=2.3cm] (own) at (5.6,0) {Owner\\ sets $B$, $V$; decides\\ escalations, top-ups};
  \node[box, draw=blue!50!black, fill=blue!6, text width=4.0cm, minimum height=1.25cm] (iss) at (2.15,-1.75) {\textbf{Elastic credential issuer} (trusted)\\ ceiling $C$; stiffness $s$; hardening $h$\\ budget $b_t$ (token bucket); leases $\ell_{\max}$};
  \node[box, text width=1.5cm] (aud) at (5.75,-1.75) {Audit of the grant log (slow loop)};
  \node[box, draw=red!50!black, fill=red!5, text width=1.65cm] (agt) at (0,-3.65) {Agent (LLM planner), untrusted};
  \node[box, text width=1.6cm] (tool) at (4.3,-3.65) {Tool server};
  \draw[rounded corners=3pt, fill=gray!10, draw=gray!60!black] (-0.95,-4.45) rectangle (6.6,-5.25);
  \node[align=center] at (2.83,-4.85) {Existing deployment: identities, roles, ACLs and resources (unchanged)};
  \draw[arr] (sec) -- (sec |- iss.north);
  \draw[arr] (own.west) -| ([xshift=30pt]iss.north) node[lab, pos=0.25, above] {$B$, $V$, top-ups};
  \draw[arr] ([xshift=-6pt]agt.north) -- ([xshift=-6pt]agt.north |- iss.south) node[lab, midway, left, align=right] {request,\\ release};
  \draw[arr, dashed] ([xshift=6pt]agt.north |- iss.south) -- ([xshift=6pt]agt.north) node[lab, midway, right, align=left] {costs, idle cost,\\ budget, reasons};
  \draw[arr] (iss.south -| tool.north) -- (tool.north) node[lab, midway, right, align=left] {short-lived,\\ scoped credential};
  \draw[arr] (agt) -- (tool) node[lab, midway, below] {tool calls};
  \draw[arr] (tool.south) -- (tool.south |- 0,-4.45);
  \draw[arr] (iss) -- (aud) node[lab, midway, above] {log};
  \draw[arr] (aud) -- (own) node[lab, midway, right, align=left] {re-baseline,\\ narrow};
  \coordinate (esc) at ([xshift=-8pt]iss.north east);
  \draw[arr, red!50!black] (esc) -- (esc |- 0,-0.8) -| ([xshift=-14pt]own.south);
  \node[lab, text=red!40!black, anchor=south] at (4.45,-0.8) {escalation};
\end{tikzpicture}
\vspace{-0.6\baselineskip}
\caption{The elasticity concept
}
\label{fig:arch}
\end{figure}

In terms of where elasticity sits, agents operate on top of conventional deployments. Beneath the agent layer, the credentials held by tool servers reach into existing identities, roles, ACLs and resources, which change slowly and are reviewed periodically, if at all. Elasticity is implemented between the two, in the component that issues credentials (Figure~\ref{fig:arch}). The planner requests and releases authority and is told its cost, while the issuer applies the ceiling, the stiffness, the budget, and the leases, and approves a short-term, scoped credential that the tool server presents to the existing deployment, whose own enforcement is unchanged.

\subsection{Threat Model}
\label{sec:threat}

Agents access resources only through credentials. The \emph{trusted} components are the credential issuer, which approves, rescopes, and revokes credentials and keeps the budget, the resource-side enforcement of scopes and ACLs, the store of baselines, envelopes, ceilings, and costs, and an append-only log of grants and calls. The \emph{potentially adversarial} components are the planner, the model and everything in its context, including task input, retrieved content, and tool descriptions~\cite{greshake2023indirect,wang2025mindguard,li2026ace}. Sub-agents, tool servers (which can exercise the full scope of the credentials they hold) and any LLM-based advisor are also treated as potentially adversarial. Owners are assumed to be benign, although they may make mistakes. The \emph{objective} is that no agent holds authority that violates a ceiling, that elevated authority is bounded over any interval regardless of the agent's behaviour, and that tasks still complete.
A compromised identity provider, misuse by those who set ceilings and costs, and credentials or channels missing from the declared deployment are out of scope. 

\subsection{Baseline, Envelope and Ceiling}
Each agent $a$ has a baseline $B(a)$, the authority it holds at rest, and an envelope $V(a) \supseteq B(a)$, set by its owner, beyond which nothing can be granted without a change of policy. It is subject to a ceiling $C$, set by the organisation's security function. Figure~\ref{fig:elastic} provides a graphical illustration of these concepts.%
Many dangerous authorities are combinations rather than single operations, so each unit carries properties, and each invariant forbids holding all properties in a set $P_i$ together. Following the Agents Rule of Two~\cite{meta2025ruleoftwo}, one invariant forbids untrusted input ($U$), sensitive data ($S$) and an external or state-changing channel ($X$) together. A single-property invariant forbids an operation outright, such as a repository deletion. Invariants are preserved when authority is removed, so recoil can never violate them and $B(a)$ must satisfy $C$. Because information read into a context outlives the authority that reads it, properties are \emph{sticky}. This means that the session property set $\Pi_t(a)$ contains the properties of every unit the agent has held since its session began, including units that have since recoiled, and a session ends only when the agent's context is discarded. This follows the rule of two, which applies within a session~\cite{meta2025ruleoftwo}. Sessions are created by the issuer when an owner initiates a task, so an agent cannot reset its properties by asking for a new one, and sub-agents inherit the properties. Sticky properties are an enforceable approximation to information-flow history, not complete information-flow control, as they assume that a session's context, memory and caches are discarded when it ends, and they cannot follow data passed to an unrelated principal. A request is checked against $C$, as a predicate over $\Pi_t(a)$ and the properties requested, before any cost is computed. Like permission boundaries~\cite{aws2026boundaries} and dynamic separation of duty~\cite{ferraiolo2001nist}, the ceiling's invariants restrict combinations that may be activated within a session. Unlike conventional dynamic separation of duty, however, the constraint is history-sensitive. Security-relevant properties remain attached to the session after the authority that introduced them has been withdrawn.

\begin{table}[tbp]
\centering
\caption{Notation used in this work. An authority unit $x$ is an operation class on a resource, and each grant $g$ confers one unit $x_g$.}
\label{tab:notation}
\footnotesize
\begin{tabular}{@{}L{1.45cm}L{6.1cm}@{}}
\toprule
Symbol & Meaning \\
\midrule
$B(a)$, $V(a)$ & Baseline (held at rest, at no cost) and envelope (the most agent $a$ may hold without the owner changing policy) \\
$H_t(a)$ & Authority held at time $t$, with $B(a) \subseteq H_t(a) \subseteq V(a)$ \\
$\mathit{grants}_t(a)$ & Elevated grants held at $t$: $H_t(a) = B(a) \cup \{x_g : g \in \mathit{grants}_t(a)\}$ \\
$C$; $P_i$ & Ceiling: invariants, each forbidding the properties $P_i$ together \\
$U$, $S$, $X$ & Properties: untrusted input, sensitive data, external or state-changing channel \\
$\Pi_t(a)$ & Session properties: of every unit held since the session began \\
$w(x)$ & Power of unit $x$, set by operation class \\
$\phi(\Pi)$, $s(\Pi)$ & Proximity of $\Pi$ to the nearest invariant; stiffness $1/(1-\phi)$ \\
$h_g$; $\lambda$, $T$ & Hardening factor of grant $g$; its weight and half-life \\
$e_t$, $\rho_t$ & Extension $\sum_g w(x_g)$; cost rate $s(\Pi_t(a)) \sum_g h_g w(x_g)$ \\
$\ell_{\max}(x)$, $L$ & Maximum lease $L/w(x)$, with lease constant $L$ \\
$b_t$, $\beta$, $r$ & Budget at $t$, its capacity and its refill rate \\
$\delta$; $\Delta_{[t_0,t_1]}$ & Admission interval; top-ups granted within $[t_0,t_1]$ \\
$\varepsilon$ & Revocation bound of the enforcement system \\
$\theta$, $n$ & Plasticity: leased in $\theta$ of the last $n$ tasks \\
\bottomrule
\end{tabular}
\end{table}

\subsection{Power, Stiffness and Cost}
\label{sec:stiffness}
The power $w(x)$ of a unit reflects what its operation class can change. Platforms already order or classify such operations. For example, GitHub's repository roles run ``from least access to most access'' through Read, Triage, Write, Maintain and Admin~\cite{github2026roles}, and AWS classifies every action as List, Read, Tagging, Write or Permissions management~\cite{aws2026accesslevels}. To convert to a numeric score, the three classes could be weighted read 1, write 3 and administer 9, where administering means changing settings or who may do what, so that each class costs up to three of the class below, following super-linear risk weights in agent tool gating~\cite{iyer2026racg}. Power deliberately does not capture confidentiality. Reading private data is a read, and its risk enters through the property $S$ and its proximity to the ceiling. Cost is therefore not an objective measure of how privileged an action is, nor a probability of compromise or an estimate of loss. Instead, it is an administratively defined resistance signal, combining authority, exposure, and security context, used to order alternatives and to measure elevated exposure.

The cost rate then depends on two quantities, namely how much authority is held above the baseline and how close the session is to violating the ceiling. The \emph{extension} of agent $a$ at time $t$ is $e_t = \sum_{g \in \mathit{grants}_t(a)} w(x_g)$, the power it has above its baseline. The \emph{stiffness} is a multiplier that grows with proximity to the ceiling. For a property set $\Pi$, the proximity to the invariant $i$ is $|P_i \cap \Pi|/|P_i|$, $\phi(\Pi)$ is the maximum over the invariants, and $s(\Pi) = 1/(1-\phi(\Pi))$, which would be infinite exactly when $\Pi$ violates $C$. Such requests are refused by the predicate check. Under the rule of two, one of the three properties gives $s = 1.5$, two give $s = 3$, and the third cannot be bought. The cost of each grant is multiplied by a hardening factor $h_g \ge 1$ (Section~\ref{sec:budget}), so the cost rate is
\begin{equation}
\rho_t \;=\; s\big(\Pi_t(a)\big) \sum_{g \in \mathit{grants}_t(a)} h_g \, w(x_g) \;\ge\; s\big(\Pi_t(a)\big)\, e_t,
\label{eq:rate}
\end{equation}
The cost is \emph{superadditive}, as adding a unit that brings the session closer to an invariant raises the cost of every elevated unit already held, so a combination costs at least as much as its parts, and strictly more when it raises proximity. The cost is also \emph{state-based}. Without hardening ($h_g = 1$), it depends only on the authority held and the property of the session, so assembling a combination close to a ceiling one small request at a time is no cheaper than asking for it at once, and hardening only makes splitting more expensive. This is the reason for measuring held authority rather than each access.

\subsection{Leases and Recoil}
Every elevation is granted as a lease rather than as standing authority. A request states a lease, which is capped for each unit at $\ell_{\max}(x) = L/w(x)$, so a read may be held for 60 minutes and a write for 20 with a lease constant $L = 60$ minutes. Authority recoils, that is, returns to the baseline, in three ways, as (1) the issuer withdraws a unit when its lease expires, (2) the planner may release a unit at any time, which stops its cost at once, and (3) the issuer withdraws all elevated units when the budget reaches zero. Recoil requires that newly acquired credentials expire no later than their lease, as short-lived and attenuated credentials allow~\cite{birgisson2014macaroons,south2025delegation}, and that earlier revocation can be checked online during use, as usage control and continuous access evaluation allow~\cite{park2004ucon,openid2025caep,salim2012thesis}. Recoil is therefore enforced by the issuer and does not depend on the agent's cooperation.

\subsection{Budget, Hardening and Escalation}
\label{sec:budget}
The costs are drawn from a budget $b_t$ that refills at a rate $r$ up to a capacity $\beta$, as a token bucket:
\begin{equation}
\dot b_t = \begin{cases} r - \rho_t, & 0 < b_t < \beta,\\ \min(0,\, r - \rho_t), & b_t \ge \beta,\\ \max(0,\, r - \rho_t), & b_t = 0,\end{cases}
\label{eq:bucket}
\end{equation}
and the issuer revokes all elevated grants when $b_t$ reaches zero, so the budget is never negative, and with $\rho_t = 0$ it then refills. Budgets and session properties attach to the owner's credential lineage, not to each agent instance, so sub-agents share those of their parent. A request is accepted only if $b_t \ge \delta \rho'$, where $\rho'$ is the cost rate after the grant, so the budget could support it for an admission interval $\delta$ without refill. Repeated requests increase cost through a \emph{hardening} factor, $h_g = 1 + \lambda \sum_j 2^{-(t_g - t_j)/T}$, which is fixed when grant $g$ is made at $t_g$, from the times $t_j < t_g$ of all previous requests in the agent's lineage, including the refused ones, and decays with half-life $T$. A burst of requests, produced by both transient failures~\cite{yang2026toolprivbench} and injected loops, therefore, becomes expensive quickly, and the effect decays once the burst ends. Repetition has been used as a risk signal before~\cite{celikel2009risk}. A request beyond the envelope or the ceiling is refused. A request within both that the budget cannot support becomes an \emph{escalation}, in which case the pending request is raised to the owner. If the owner approves, the issuer adds a logged top-up equal to the cost of the approved requests over their leases. A top-up is restricted credit. It is counted in $b_t$, can be taken above $\beta$, can be spent only on approved units and resources in the same session and task, and expires when approved leases end.

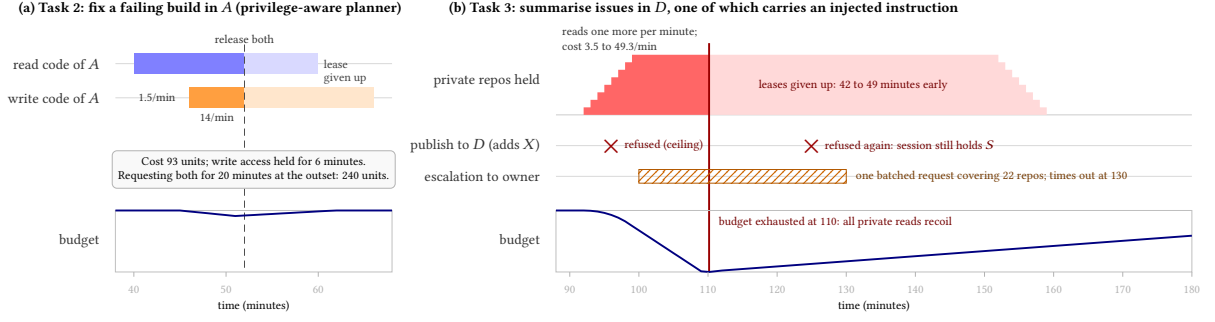
\begin{figure*}[tbp]
\centering
\renewcommand{\scriptsize}{\fontsize{6}{7}\selectfont}\renewcommand{\tiny}{\fontsize{5}{6}\selectfont}
\resizebox{\textwidth}{!}{%
\begin{tikzpicture}[font=\scriptsize,
  lane/.style={anchor=east, font=\scriptsize, align=right, text=black!80},
  ann/.style={font=\tiny, text=black!80, align=left, inner sep=1pt},
  ttl/.style={anchor=west, font=\scriptsize\bfseries}]
  \node[ttl] at (0,0.4) {(a) Task 2: fix a failing build in $A$ (privilege-aware planner)};
  \node[lane] at (1.45,-0.4) {read code of $A$};
  \node[lane] at (1.45,-0.9) {write code of $A$};
  \node[lane] at (1.45,-3.0) {budget};
  \draw[gray!30] (1.550,-0.400) -- (5.600,-0.400); \draw[gray!30] (1.550,-0.900) -- (5.600,-0.900);
  \fill[blue!50] (1.820,-0.550) rectangle (3.440,-0.250);
  \fill[blue!15] (3.440,-0.550) rectangle (4.520,-0.250);
  \fill[orange!75] (2.630,-1.050) rectangle (3.440,-0.750);
  \fill[orange!20] (3.440,-1.050) rectangle (5.330,-0.750);
  \node[ann, anchor=west] at (1.847,-0.900) {1.5/min};
  \node[ann, anchor=north] at (3.035,-1.080) {14/min};
  \draw[gray!60!black, densely dashed] (3.440,-0.120) -- (3.440,-3.450);
  \node[ann, anchor=south] at (3.440,-0.140) {release both};
  \node[ann, anchor=west, text=gray!50!black] at (4.561,-0.400) {lease};
  \node[ann, anchor=west, text=gray!50!black] at (4.561,-0.620) {given up};
  \node[draw=gray!60, rounded corners=2pt, fill=gray!5, font=\tiny, align=center, inner sep=2.5pt] at (3.575,-1.950) {Cost 93 units; write access held for 6 minutes.\\ Requesting both for 20 minutes at the outset: 240 units.};
  \draw[gray!50] (1.550,-3.450) rectangle (5.600,-2.550);
  \draw[thick, blue!50!black] plot coordinates {(1.550,-2.550) (1.685,-2.550) (1.820,-2.550) (1.955,-2.550) (2.090,-2.550) (2.225,-2.550) (2.360,-2.550) (2.495,-2.550) (2.630,-2.564) (2.765,-2.577) (2.900,-2.590) (3.035,-2.604) (3.170,-2.617) (3.305,-2.631) (3.440,-2.623) (3.575,-2.616) (3.710,-2.608) (3.845,-2.601) (3.980,-2.593) (4.115,-2.586) (4.250,-2.578) (4.385,-2.571) (4.520,-2.563) (4.655,-2.556) (4.790,-2.550) (4.925,-2.550) (5.060,-2.550) (5.195,-2.550) (5.330,-2.550) (5.465,-2.550) (5.600,-2.550)};
  \draw[gray!60] (1.820,-3.450) -- (1.820,-3.520) node[below, font=\tiny] {40};
  \draw[gray!60] (3.170,-3.450) -- (3.170,-3.520) node[below, font=\tiny] {50};
  \draw[gray!60] (4.520,-3.450) -- (4.520,-3.520) node[below, font=\tiny] {60};
  \node[font=\tiny] at (3.575,-3.950) {time (minutes)};
  \node[ttl] at (6.3,0.4) {(b) Task 3: summarise issues in $D$, one of which carries an injected instruction};
  \node[lane] at (7.9,-0.65) {private repos held};
  \node[lane] at (7.9,-1.6) {publish to $D$ (adds $X$)};
  \node[lane] at (7.9,-2.05) {escalation to owner};
  \node[lane] at (7.9,-3.0) {budget};
  \draw[gray!30] (8.000,-1.600) -- (17.310,-1.600); \draw[gray!30] (8.000,-2.050) -- (17.310,-2.050);
  \draw[gray!40] (8.000,-1.150) -- (17.310,-1.150);
  \fill[red!60] (8.405,-1.150) -- (8.405,-1.150) -- (8.405,-1.040) -- (8.506,-1.040) -- (8.506,-0.930) -- (8.607,-0.930) -- (8.607,-0.820) -- (8.708,-0.820) -- (8.708,-0.710) -- (8.810,-0.710) -- (8.810,-0.600) -- (8.911,-0.600) -- (8.911,-0.490) -- (9.012,-0.490) -- (9.012,-0.380) -- (9.113,-0.380) -- (9.113,-0.270) -- (10.242,-0.270) -- (10.242,-1.150) -- cycle;
  \fill[red!15] (10.242,-1.150) -- (10.242,-0.270) -- (14.477,-0.270) -- (14.477,-0.380) -- (14.578,-0.380) -- (14.578,-0.490) -- (14.679,-0.490) -- (14.679,-0.600) -- (14.780,-0.600) -- (14.780,-0.710) -- (14.882,-0.710) -- (14.882,-0.820) -- (14.983,-0.820) -- (14.983,-0.930) -- (15.084,-0.930) -- (15.084,-1.040) -- (15.185,-1.040) -- (15.185,-1.150) -- cycle;
  \node[ann, anchor=south west] at (8.051,-0.070) {reads one more per minute;};
  \node[ann, anchor=south west] at (8.051,-0.250) {cost 3.5 to 49.3/min};
  \node[ann, anchor=center, text=red!45!black] at (12.352,-0.710) {leases given up: 42 to 49 minutes early};
  \draw[red!60!black, thick] (10.242,-0.100) -- (10.242,-3.450);
  \node[text=red!60!black, font=\bfseries] at (8.810,-1.600) {$\times$};
  \node[text=red!60!black, font=\bfseries] at (11.744,-1.600) {$\times$};
  \node[ann, anchor=west, text=red!45!black] at (8.960,-1.600) {refused (ceiling)};
  \node[ann, anchor=west, text=red!45!black] at (11.894,-1.600) {refused again: session still holds $S$};
  \fill[pattern=north east lines, pattern color=orange!80!black] (9.214,-2.150) rectangle (12.250,-1.950);
  \draw[orange!70!black] (9.214,-2.150) rectangle (12.250,-1.950);
  \node[ann, anchor=west, text=orange!45!black] at (12.350,-2.050) {one batched request covering 22 repos; times out at 130};
  \draw[gray!50] (8.000,-3.450) rectangle (17.310,-2.550);
  \draw[thick, blue!50!black] plot coordinates {(8.000,-2.550) (8.101,-2.550) (8.202,-2.550) (8.304,-2.550) (8.405,-2.550) (8.506,-2.554) (8.607,-2.565) (8.708,-2.585) (8.810,-2.615) (8.911,-2.655) (9.012,-2.708) (9.113,-2.775) (9.214,-2.841) (9.316,-2.908) (9.417,-2.974) (9.518,-3.041) (9.619,-3.107) (9.720,-3.174) (9.822,-3.240) (9.923,-3.307) (10.024,-3.373) (10.125,-3.440) (10.242,-3.450) (10.429,-3.429) (10.631,-3.414) (10.834,-3.399) (11.036,-3.384) (11.238,-3.369) (11.441,-3.354) (11.643,-3.339) (11.846,-3.324) (12.048,-3.309) (12.250,-3.294) (12.453,-3.279) (12.655,-3.264) (12.858,-3.249) (13.060,-3.234) (13.262,-3.219) (13.465,-3.204) (13.667,-3.189) (13.870,-3.174) (14.072,-3.159) (14.274,-3.144) (14.477,-3.129) (14.679,-3.114) (14.882,-3.099) (15.084,-3.084) (15.286,-3.069) (15.489,-3.054) (15.691,-3.039) (15.894,-3.024) (16.096,-3.009) (16.298,-2.994) (16.501,-2.979) (16.703,-2.964) (16.906,-2.949) (17.108,-2.934) (17.310,-2.919)};
  \node[ann, anchor=west, text=red!45!black] at (10.342,-2.720) {budget exhausted at 110: all private reads recoil};
  \draw[gray!60] (8.202,-3.450) -- (8.202,-3.520) node[below, font=\tiny] {90};
  \draw[gray!60] (9.214,-3.450) -- (9.214,-3.520) node[below, font=\tiny] {100};
  \draw[gray!60] (10.226,-3.450) -- (10.226,-3.520) node[below, font=\tiny] {110};
  \draw[gray!60] (11.238,-3.450) -- (11.238,-3.520) node[below, font=\tiny] {120};
  \draw[gray!60] (12.250,-3.450) -- (12.250,-3.520) node[below, font=\tiny] {130};
  \draw[gray!60] (13.262,-3.450) -- (13.262,-3.520) node[below, font=\tiny] {140};
  \draw[gray!60] (14.274,-3.450) -- (14.274,-3.520) node[below, font=\tiny] {150};
  \draw[gray!60] (15.286,-3.450) -- (15.286,-3.520) node[below, font=\tiny] {160};
  \draw[gray!60] (16.298,-3.450) -- (16.298,-3.520) node[below, font=\tiny] {170};
  \draw[gray!60] (17.310,-3.450) -- (17.310,-3.520) node[below, font=\tiny] {180};
  \node[font=\tiny] at (12.655,-3.950) {time (minutes)};
\end{tikzpicture}}
\caption{The running example under elastic least privilege. Solid bars and areas show elevated authority held, pale extensions show lease time given up, and the lower panels show the budget, from 0 to 600 units. Task 1 uses only the baseline and is omitted. (a) A privilege-aware planner reads before writing and releases both units when done. (b) The ceiling refuses exfiltration while private data is held and after it has recoiled, and the budget bounds how much private code the injected agent can hold, and for how long.}
\label{fig:trace}
\end{figure*}

\subsection{Privilege Information for the Planner}
\label{sec:visible}
In most current agent deployments, privilege is enforced when a request is made but is not presented as something the planner can compare in advance. Elastic least privilege makes privilege, and over-privilege, part of what the planner reasons about, through four signals that the issuer computes and the tool interface shows.
\begin{itemize}
  \item \textbf{Marginal cost.} For each candidate tool, the cost rate its authority would add, given what is already held, so that a read-only tool and one needing write access are distinguishable before either is chosen.
  \item \textbf{Idle cost.} What elevated units held but not exercised are costing, so that over-privilege becomes a visible running cost.
  \item \textbf{Release.} The planner may return any elevated unit early, which stops its cost at once. Because an exhausted budget stops the task, shedding over-privilege serves the planner's own objective.
  \item \textbf{Reasons.} A refusal states its cause (ceiling, envelope or budget) and any lower-cost alternative.
\end{itemize}
Budget-aware agents already condition their behaviour on a remaining budget~\cite{liu2025budgetaware}. Costs are computed outside the model, as in the Irreversibility Budget~\cite{mohammadi2026irreversibility}, but shown to it, since misusing them cannot weaken any of the guarantees in Section~\ref{sec:guarantees}.

\subsection{Plasticity: Re-baselining Across Tasks}
Within a task, authority changes through leases and the budget. Across tasks, a slower governance process, which we call \emph{plasticity}, can change the baseline itself. An audit of the grant log counts recurring elevations. If the same unit is leased in at least $\theta$ of the agent's last $n$ tasks, the audit proposes to the owner that it be added to $B(a)$. Conversely, baseline units that are not exercised over a review window are proposed for removal, as cloud right-sizing already does for unused access~\cite{barr2023unusedaccess}. Unlike B-RBAC's budget recalibration~\cite{salim2013brbac}, plasticity moves the grant itself, so that recurring needs stop incurring a cost while permission creep requires an owner's decision. Recurrent use is evidence of operational need, not evidence that holding the authority permanently is safe, so a proposal starts a review rather than presuming inclusion. More specifically, a proposed baseline is checked against $C$ and against a standing-authority policy, which may forbid holding permanently what is acceptable briefly. 

\subsection{Enforced Guarantees}
\label{sec:guarantees}
In this section, we state what under the following assumptions, the issuer enforces regardless of how the planner behaves. The guarantees depend on four assumptions, which are that (1) every route by which the agent can act, including tool servers and the credentials they hold, passes through the issuer, which we term \emph{coverage soundness}, (2) the baseline itself satisfies the ceiling, (3) properties such as $U$, $S$ and $X$ are labelled correctly, and (4) credentials expire with their lease and can be revoked within a bound $\varepsilon$. The first assumption is the most demanding. More specifically, the issuer can only measure the authority it has issued, whereas the security objective concerns everything the agent can actually cause. Ambient credentials inside tools, tokens retained by tool servers and confused deputies all widen this gap, which is why sound accounting for authority is itself a research question. Under these assumptions, four properties hold for each agent and any sub-agents it delegates to.

\begin{itemize}
  \item \textbf{P1 (ceiling).} $\Pi_t(a)$ satisfies $C$ at every $t$, because the ceiling is checked as a predicate on the session's sticky properties before any cost is computed, so no budget or top-up can admit a violating request, and recoil only removes authority.
  \item \textbf{P2 (bounded elevated exposure).} Over any interval $[t_0, t_1]$, $\int_{t_0}^{t_1} e_t\, \mathrm{d}t \le b_{t_0} + r\,(t_1 - t_0) + \Delta_{[t_0,t_1]}$, where $\Delta_{[t_0,t_1]}$ is the sum of top-ups granted within the interval. This follows from Equation~\ref{eq:bucket}, since refill adds at most $r$ per unit time, the budget is never negative, and it is charged at least $e_t$ per unit time, since $s, h_g \ge 1$. If no top-up credit is unexpired at $t_0$, then $b_{t_0} \le \beta$.
  \item \textbf{P3 (recoil).} Without new grants, $H_t(a)$ returns to $B(a)$ within the longest remaining lease of any elevated grant, plus $\varepsilon$, and within $\varepsilon$ of the budget being exhausted.
  \item \textbf{P4 (independence).} P1 to P3 hold whatever the model does, whether or not the planner reads its costs, and whatever any LLM-based advisor recommends.
\end{itemize}
In practice, revocation is not instantaneous. It takes time to propagate, tokens may be cached, and requests already in flight complete under the revoked credential. Therefore, the exposure limit may be exceeded by the maximum authority held during the revocation delay $\varepsilon$, which a prototype would need to measure. It is worth noting that this bound is the budget bound of B-RBAC~\cite{salim2013brbac} restated for authority held over time, and that it holds for any metered budget. Stiffness, hardening, and visible costs therefore add no guarantee of their own, and their role is to shape what a bounded budget buys and to tell the planner what it is buying. Finally, the bound limits exposure rather than harm. A disclosure can happen in seconds, so authority whose misuse cannot be undone should be placed in the ceiling or require per-use approval.

\section{Worked Example}
\label{sec:example}

This section applies the model to the running coding-agent example. The parameters are illustrative and were selected to make the model's behaviour visible. Figure~\ref{fig:trace} was generated by a short simulation and should not be interpreted as an empirical result.

In the configuration used for the example, the user maintains a public repository $A$, contributes to a public repository $D$, and owns 48 private repositories. Each task runs in a new session. The baseline is to read issues on $A$ and $D$, which carries $U$, because public issues are untrusted input. The envelope allows for reading and writing the contents of all 50 repositories and opening pull requests. The ceiling forbids deletion, administrative changes, and the rule-of-two combination. Private contents carry $S$, and writes to public repositories carry $X$. The budget has capacity $\beta = 600$ units and refills at $r = 5$ units per minute. Hardening uses $\lambda = 0.25$ and $T = 30$ minutes, the admission interval is $\delta = 10$ minutes, and escalations time out after 30 minutes.

Task 2 is a privilege-aware build fix. At minute 40, the agent is asked to fix a failing build in $A$. Its tool interface shows that reading $A$'s code would cost 1.5 units per minute, and reading and writing it 12. In an illustrative privilege-aware trajectory, the planner first reads, to diagnose the failure. At minute 46 it requests write access, whose cost rises to 14 units per minute with slight hardening, and at minute 52, once the fix is pushed, it releases both units. The task cost 93 units and held write access for 6 minutes. Requesting read and write access for 20 minutes at the outset would have cost 240 units and held write access more than three times as long.

Task 3 is a prompt-injection attack. At minute 90, the agent summarises issues in $D$, one of which carries an injected instruction to collect the user's private code and publish it~\cite{milanta2025githubmcp}. From minute 92, it requests read access to one private repository per minute. Each holds $U$ and $S$, so $s = 3$, and hardening raises $h_g$ from 1.16 for the first request to 2.97 for the eighth, so the cost rate climbs from 3.5 to 49.3 units per minute. At minute 96, it requests a pull request on $D$, which would add $X$, and the ceiling refuses it regardless of budget. At minute 100, the ninth repository is unaffordable, and a single escalation, to which the next 21 requests are added, shows the owner one anomalous request for 22 private repositories during an issue summary. Just after minute 110, the budget is exhausted, and all eight elevated reads recoil, between 42 and 49 minutes before their leases would have expired. At minute 125, a second pull request is refused, because the session still carries $S$, and at minute 130 the escalation times out.

Comparing the two tasks, if the planner responds to its costs in Task 2, least privilege is also the cheaper plan, whereas a conventional fixed grant or undifferentiated approval prompt would not ordinarily distinguish these alternatives. Whether planners do respond is an empirical question. In Task 3, the ceiling, not the budget, prevented the exfiltration, both while the reads were held and after they recoiled. This holds at any cost because the properties are sticky. The budget limited the injected agent to holding eight private repositories for 11 to 18 minutes each. Hardening made the burst increasingly costly, and batching gave the owner one decision rather than 22.

\section{Design Choices and Limitations}
\label{sec:limits}

The concept leaves several questions unresolved, and we summarise the largest below.

\begin{itemize}
  \item \textbf{Costs are arbitrary.} Their absolute values are, but their ordering need not be. P1 to P4 hold for any positive weights. Calibration changes the frequency with which tasks reach the owner and the amount of exposure a budget buys.
  \item \textbf{Budget draining.} An injected agent can spend the budget and block legitimate work. This is an availability attack, not a safety failure, and exhaustion during an unrelated task is itself a finding for audit~\cite{salim2013brbac}.
  \item \textbf{Liveness.} Too small a budget makes the agent useless, and too large a budget removes the resistance. The critique that budgets discourage necessary access~\cite{liu2011incentive} is answered by escalation rather than denial.
  \item \textbf{Sticky properties constrain legitimate work.}  Once a session combines untrusted input with private data, it cannot write publicly. A legitimate workflow such as porting private code to a public repository therefore requires an owner-approved per-use exception or a trusted declassification and hand-off mechanism that produces sanitised input for a separate session. Merely beginning a new session is insufficient if that session reacquires the same security-relevant properties.
  \item \textbf{Visible costs reveal sensitivity.} A cost can show an injected planner which resources carry $S$, as B-RBAC's authors noted that a cost ``may reveal some information about the importance of the resource''~\cite{salim2013brbac}, although coarse bands may help (R6).
\end{itemize}

\section{Open Research Agenda}
\label{sec:agenda}

The concept presents a promising approach to agent permissioning. However, it has limitations and raises seven research questions, which together form a research agenda.

\paragraph{R1: Privilege-aware planning.} \emph{Does making privilege and over-privilege visible change what agents do?} This is the central hypothesis of the paper, and it could fail in two ways, as planners may ignore the costs they are shown or minimise them at the expense of completing the task. A first study could compare four conditions on ToolPrivBench~\cite{yang2026toolprivbench} and AgentDojo~\cite{debenedetti2024agentdojo}, namely (1) no signal, (2) a prompt instruction to prefer least privilege, (3) visible costs without enforcement, and (4) visible costs with an enforced budget. It would measure over-privileged tool selection, task completion and authority-time, the elevated authority held integrated over a task. More specifically, comparing the third and fourth conditions separates the effect of information from that of constraint. If visible costs alone reduce over-privilege, steering is inexpensive to deploy, whereas if only an enforced budget does, the value of the approach lies in enforcement. A second study could use costs as a reward in privilege-aware post-training~\cite{yang2026toolprivbench}, testing whether the preference transfers to unseen tools. A negative result would still show that privilege must be enforced rather than communicated.

\paragraph{R2: Calibrating power and stiffness.} \emph{Which functions make agents prefer lower privilege without blocking tasks?} Power classes can follow platform orderings~\cite{github2026roles,aws2026accesslevels}, with parameters fitted by replaying recorded trajectories. Given a corpus of benign and adversarial traces, the calibration problem is to choose the power weights, refill rate and hardening parameters so that benign tasks rarely reach the owner while injected bursts exhaust the budget early. This is a multi-objective problem, and presenting owners with candidate configurations along the trade-off between friction and exposure would let each organisation choose its own position. It is worth noting that the enforced guarantees hold for any positive weights, so calibration determines usability and the exposure a budget buys, not whether the ceiling holds.

\paragraph{R3: Labelling properties soundly.} \emph{How can properties such as $U$, $S$ and $X$ be assigned to tools described in natural language?} Labels are safety-critical, since the ceiling is only as sound as the properties over which it is evaluated. For example, a tool wrongly labelled as having no external channel would permit the rule-of-two combination. An LLM could propose labels from tool descriptions, schemas and code, each then validated deterministically against the tool's declared scopes and, where source code is available, by program analysis of whether network or write operations are reachable. This generate-and-check pattern has been applied to LLM-generated code repair, where a bounded model checker validates each candidate fix and returns a counterexample when it fails~\cite{tihanyi2023selfhealing}, which here would identify the operation contradicting a label. Vague properties such as ``sensitive'' suit graded membership, as in risk-weighted analysis of over-entitlement using fuzzy logic~\cite{parkinson2022fuzzy}.

\paragraph{R4: A verified elastic credential issuer.} \emph{Can the issuer be built on existing mechanisms, at acceptable cost, and shown to enforce its guarantees?} A GitHub App can already grant installation tokens that expire after one hour and are restricted to named repositories and permissions~\cite{github2026apptokens}, although shorter leases need online revocation or a mediating proxy. A prototype integrated with MCP authorisation~\cite{mcp2025authorization} would measure latency, the revocation bound $\varepsilon$, and the gap between declared and effective authority~\cite{li2025mcpprivilege}. The issuer is also small, trusted and precisely specified, which makes it suited to formal verification. Its budget, leases, sticky properties and admission check form a state machine whose guarantees are invariants over sequences of requests, releases, expiries and top-ups. Bounded model checking of the implementation~\cite{gadelha2018esbmc}, including an issuer written in Python~\cite{farias2024esbmcpython}, could check that the ceiling holds in every reachable state, that elevated exposure never exceeds the budget bound, and that authority recoils within the longest lease, returning a counterexample trace whenever a property fails. As the exposure bound is stated over continuous time, verification would require a discretised form whose error must itself be bounded.

\paragraph{R5: Owners, escalation and plasticity.} \emph{How should escalations be batched and explained, and when should recurring demand move the baseline?} User studies could compare batched, explained escalations with individual prompts on injected requests~\cite{hughes2026automode,haselmehri2025explainable}, measuring both how often owners approve malicious requests and how long decisions take. The explanation shown to an owner should be constructed from trusted facts held by the issuer, such as the task, the units requested and the session's properties, rather than from the agent's own account, which an injected instruction controls. Replay of grant logs would set $\theta$ and $n$, trading permission creep against friction. Owners must also review re-baselining proposals over time, since individually reasonable additions can accumulate into over-entitlement.

\paragraph{R6: Attacks on elasticity.} \emph{Can an adversary exploit budgets, hardening, sessions, or escalation?} Attacks include draining the budget, crafting routine-looking escalations, splitting a combination into many small requests, and using costs to locate sensitive resources. Beyond red-team studies in AgentDojo~\cite{debenedetti2024agentdojo}, the issuer could be model checked against an adversarial agent that may issue any sequence of requests. Rather than sampling attacks, this would explore every request sequence up to a bound and either confirm that the guarantees hold or produce the shortest sequence that violates them. Such counterexamples would also inform calibration, for example by revealing the request pattern that maximises the private data held before the budget is exhausted.

\paragraph{R7: Beyond one agent.} \emph{How should elasticity compose across delegation chains?} When agents delegate to sub-agents or collaborate across owners, budgets and session properties must be shared or partitioned without allowing a chain to hold more than any of its members could alone. Credentials that can only be attenuated as they are delegated~\cite{birgisson2014macaroons} offer one basis, but the cost of authority held by a sub-agent must still be charged to the lineage that requested it. Combining elasticity with information-flow defences~\cite{debenedetti2025camel,costa2025fides} is a natural next step, since sticky properties approximate the information-flow history that those defences track more precisely.

The vision as a whole could be evaluated through replay and red-team studies, which would measure its cost (tasks completed within budget, escalations) and benefit (authority an injected agent can hold), using a shared, privilege-labelled corpus of agent traces.

\section{Conclusion}
\label{sec:conclusion}

Least privilege cannot be realised for general-purpose agents through a single static permission set, and an agent's planner is rarely shown the authority its choices require. Building on priced and budgeted access control, we have proposed \emph{elastic least privilege}, in which an agent holds a small baseline and acquires leased authority within an owner-defined envelope, at a cost that rises with power, proximity to a mandatory invariant, and repeated requests. Authority returns to the baseline when leases expire, when it is released, or when the budget is exhausted. Visible costs enable decisions that hold less authority for less time. Whether visible costs reduce over-privilege without preventing legitimate work, whether effective authority can be accounted for soundly, and whether owners can govern escalations at scale are empirical questions for the access-control community. The aim is not to replace the judgement of policy owners, but to ensure that agents exercise no more authority, for no longer, than their tasks require.

\FloatBarrier
\bibliographystyle{plainnat}
\bibliography{refs}

@article{parkinson2017ransomware,
  author  = {Simon Parkinson},
  title   = {Use of Access Control to Minimise Ransomware Impact},
  journal = {Network Security},
  volume  = {2017},
  number  = {7},
  pages   = {5--8},
  year    = {2017}
}

@article{parkinson2022fuzzy,
  author  = {Simon Parkinson and Saad Khan},
  title   = {Identifying High-Risk Over-Entitlement in Access Control Policies Using Fuzzy Logic},
  journal = {Cybersecurity},
  volume  = {5},
  number  = {6},
  year    = {2022},
  doi     = {10.1186/s42400-022-00112-1}
}

@article{saltzer1975protection,
  author  = {Jerome H. Saltzer and Michael D. Schroeder},
  title   = {The Protection of Information in Computer Systems},
  journal = {Proceedings of the IEEE},
  volume  = {63},
  number  = {9},
  pages   = {1278--1308},
  year    = {1975},
  doi     = {10.1109/PROC.1975.9939}
}

@article{ferraiolo2001nist,
  author  = {David F. Ferraiolo and Ravi Sandhu and Serban Gavrila and D. Richard Kuhn and Ramaswamy Chandramouli},
  title   = {Proposed {NIST} Standard for Role-Based Access Control},
  journal = {ACM Transactions on Information and System Security},
  volume  = {4},
  number  = {3},
  pages   = {224--274},
  year    = {2001},
  doi     = {10.1145/501978.501980}
}

@misc{mcp2025authorization,
  author       = {{Model Context Protocol}},
  title        = {Model Context Protocol Specification, Version 2026-07-28: Authorization},
  howpublished = {\url{https://modelcontextprotocol.io/specification/2026-07-28/basic/authorization}},
  year         = {2026},
  note         = {Accessed October 2026}
}

@misc{microsoft2026leastprivilege,
  author       = {{Microsoft Security}},
  title        = {Least Privilege for {AI} Agents: Identity, Access, and Tool Binding},
  howpublished = {\url{https://www.microsoft.com/en-us/security/blog/2026/07/16/least-privilege-for-ai-agents-identity-access-and-tool-binding/}},
  year         = {2026},
  note         = {Microsoft Security Blog, 16 July 2026}
}

@misc{willison2025trifecta,
  author       = {Simon Willison},
  title        = {The Lethal Trifecta for {AI} Agents: Private Data, Untrusted Content, and External Communication},
  howpublished = {\url{https://simonwillison.net/2025/Jun/16/the-lethal-trifecta/}},
  year         = {2025},
  note         = {16 June 2025}
}

@misc{shi2025progent,
  author       = {Tianneng Shi and Jingxuan He and Zhun Wang and Hongwei Li and Linyu Wu and Wenbo Guo and Dawn Song},
  title        = {Progent: Securing {AI} Agents with Privilege Control},
  year         = {2025},
  eprint       = {2504.11703},
  archiveprefix= {arXiv},
  primaryclass = {cs.CR},
  note         = {Version 3, 14 May 2026}
}

@misc{debenedetti2025camel,
  author       = {Edoardo Debenedetti and Ilia Shumailov and Tianqi Fan and Jamie Hayes and Nicholas Carlini and Daniel Fabian and Christoph Kern and Chongyang Shi and Andreas Terzis and Florian Tram{\`e}r},
  title        = {Defeating Prompt Injections by Design},
  year         = {2025},
  eprint       = {2503.18813},
  archiveprefix= {arXiv},
  primaryclass = {cs.CR}
}

@misc{costa2025fides,
  author       = {Manuel Costa and Boris K{\"o}pf and Aashish Kolluri and Andrew Paverd and Mark Russinovich and Ahmed Salem and Shruti Tople and Lukas Wutschitz and Santiago Zanella-B{\'e}guelin},
  title        = {Securing {AI} Agents with Information-Flow Control},
  year         = {2025},
  eprint       = {2505.23643},
  archiveprefix= {arXiv},
  primaryclass = {cs.CR}
}

@inproceedings{li2026ace,
  author    = {Evan Li and Tushin Mallick and Evan Rose and William Robertson and Alina Oprea and Cristina Nita-Rotaru},
  title     = {{ACE}: A Security Architecture for {LLM}-Integrated App Systems},
  booktitle = {Proceedings of the Network and Distributed System Security Symposium (NDSS)},
  year      = {2026},
  doi       = {10.14722/ndss.2026.230352}
}

@misc{li2025mcpprivilege,
  author       = {Zhihao Li and Kun Li and Boyang Ma and Minghui Xu and Yue Zhang and Xiuzhen Cheng},
  title        = {We Urgently Need Privilege Management in {MCP}: A Measurement of {API} Usage in {MCP} Ecosystems},
  year         = {2025},
  eprint       = {2507.06250},
  archiveprefix= {arXiv},
  primaryclass = {cs.CR}
}

@misc{chen2026rethinkingmcp,
  author       = {Pei Chen and Baichao An and Mengying Wu and Binwang Wan and Geng Hong and Jinsong Chen and Xudong Pan and Jiarun Dai and Min Yang},
  title        = {Rethinking {MCP} Security: A Large-Scale Study of Runtime {MCP} Servers and Security Scanner Reliability},
  year         = {2026},
  eprint       = {2607.11086},
  archiveprefix= {arXiv},
  primaryclass = {cs.CR}
}

@misc{yang2026toolprivbench,
  author       = {Kaiyue Yang and Yuyan Bu and Jingwei Yi and Yuchi Wang and Biyu Zhou and Juntao Dai and Songlin Hu and Yaodong Yang},
  title        = {When Lower Privileges Suffice: Investigating Over-Privileged Tool Selection in {LLM} Agents},
  year         = {2026},
  eprint       = {2606.20023},
  archiveprefix= {arXiv}
}

@misc{debenedetti2024agentdojo,
  author       = {Edoardo Debenedetti and Jie Zhang and Mislav Balunovi{\'c} and Luca Beurer-Kellner and Marc Fischer and Florian Tram{\`e}r},
  title        = {{AgentDojo}: A Dynamic Environment to Evaluate Prompt Injection Attacks and Defenses for {LLM} Agents},
  year         = {2024},
  eprint       = {2406.13352},
  archiveprefix= {arXiv},
  primaryclass = {cs.CR}
}

@inproceedings{greshake2023indirect,
  author    = {Kai Greshake and Sahar Abdelnabi and Shailesh Mishra and Christoph Endres and Thorsten Holz and Mario Fritz},
  title     = {Not What You've Signed Up For: Compromising Real-World {LLM}-Integrated Applications with Indirect Prompt Injection},
  booktitle = {Proceedings of the 16th ACM Workshop on Artificial Intelligence and Security (AISec)},
  year      = {2023},
  doi       = {10.1145/3605764.3623985}
}

@misc{wang2025mindguard,
  author       = {Zhiqiang Wang and Haohua Du and Guanquan Shi and Junyang Zhang and Haoran Cheng and Yunhao Yao and Kaiwen Guo and Xiang-Yang Li},
  title        = {{MindGuard}: Intrinsic Decision Inspection for Securing {LLM} Agents Against Metadata Poisoning},
  year         = {2025},
  eprint       = {2508.20412},
  archiveprefix= {arXiv},
  primaryclass = {cs.CR}
}

@inproceedings{haselmehri2025explainable,
  author    = {Gelareh Hasel Mehri and Charles Morisset and Nicola Zannone},
  title     = {Towards Explainable Access Control},
  booktitle = {Proceedings of the 30th ACM Symposium on Access Control Models and Technologies (SACMAT)},
  pages     = {117--126},
  year      = {2025},
  note      = {BlueSky paper},
  doi       = {10.1145/3734436.3734439}
}

@misc{chik2023entra,
  author       = {Joy Chik},
  title        = {New {Microsoft Entra} Features Strengthen Identity Security},
  howpublished = {Microsoft Security Blog, \url{https://www.microsoft.com/en-us/security/blog/2023/03/30/latest-microsoft-entra-advancements-strengthen-identity-security/}},
  year         = {2023},
  note         = {30 March 2023. Reports findings of the 2023 State of Cloud Permissions Risks Report}
}

@inproceedings{koishybayev2022github,
  author    = {Igibek Koishybayev and Aleksandr Nahapetyan and Raima Zachariah and Siddharth Muralee and Bradley Reaves and Alexandros Kapravelos and Aravind Machiry},
  title     = {Characterizing the Security of {GitHub} {CI} Workflows},
  booktitle = {Proceedings of the 31st USENIX Security Symposium},
  pages     = {2747--2763},
  year      = {2022}
}

@misc{barr2023unusedaccess,
  author       = {Jeff Barr},
  title        = {{IAM Access Analyzer} Updates: Find Unused Access, Check Policies before Deployment},
  howpublished = {AWS News Blog, \url{https://aws.amazon.com/blogs/aws/iam-access-analyzer-updates-find-unused-access-check-policies-before-deployment/}},
  year         = {2023},
  note         = {26 November 2023}
}

@misc{openid2025caep,
  author       = {Tim Cappalli and Atul Tulshibagwale},
  title        = {{OpenID} Continuous Access Evaluation Profile 1.0},
  howpublished = {OpenID Foundation Final Specification, \url{https://openid.net/specs/openid-caep-1_0-final.html}},
  year         = {2025},
  note         = {29 August 2025}
}

@inproceedings{south2025delegation,
  author    = {Tobin South and Samuele Marro and Thomas Hardjono and Robert Mahari and Cedric Deslandes Whitney and Alan Chan and Alex Pentland},
  title     = {Position: {AI} Agents Need Authenticated Delegation},
  booktitle = {Proceedings of the 42nd International Conference on Machine Learning (ICML)},
  series    = {Proceedings of Machine Learning Research},
  volume    = {267},
  pages     = {82211--82231},
  year      = {2025}
}

@inproceedings{terekhov2026adaptive,
  author    = {Mikhail Terekhov and Alexander Panfilov and Daniil Dzenhaliou and Caglar Gulcehre and Maksym Andriushchenko and Ameya Prabhu and Jonas Geiping},
  title     = {Adaptive Attacks on Trusted Monitors Subvert {AI} Control Protocols},
  booktitle = {International Conference on Learning Representations (ICLR)},
  year      = {2026},
  eprint    = {2510.09462},
  archiveprefix = {arXiv}
}

@misc{hughes2026automode,
  author       = {John Hughes},
  title        = {How We Built {Claude Code} Auto Mode: A Safer Way to Skip Permissions},
  howpublished = {Anthropic Engineering Blog, \url{https://www.anthropic.com/engineering/claude-code-auto-mode}},
  year         = {2026},
  note         = {25 March 2026}
}

@misc{milanta2025githubmcp,
  author       = {Marco Milanta and Luca Beurer-Kellner},
  title        = {{GitHub MCP} Exploited: Accessing Private Repositories via {MCP}},
  howpublished = {Invariant Labs blog, \url{https://invariantlabs.ai/blog/mcp-github-vulnerability}},
  year         = {2025},
  note         = {26 May 2025}
}

@misc{meta2025ruleoftwo,
  author       = {{Meta AI}},
  title        = {Agents Rule of Two: A Practical Approach to {AI} Agent Security},
  howpublished = {Meta AI blog, \url{https://ai.meta.com/blog/practical-ai-agent-security/}},
  year         = {2025},
  note         = {31 October 2025}
}

@misc{aws2026boundaries,
  author       = {{Amazon Web Services}},
  title        = {Permissions Boundaries for {IAM} Entities},
  howpublished = {AWS Identity and Access Management User Guide, \url{https://docs.aws.amazon.com/IAM/latest/UserGuide/access_policies_boundaries.html}},
  year         = {2026},
  note         = {Accessed October 2026}
}

@misc{microsoft2026pim,
  author       = {{Microsoft}},
  title        = {Assign {Microsoft Entra} Roles in Privileged Identity Management},
  howpublished = {Microsoft Learn, \url{https://learn.microsoft.com/en-us/entra/id-governance/privileged-identity-management/pim-how-to-add-role-to-user}},
  year         = {2026},
  note         = {Accessed October 2026}
}

@misc{github2026mcpserver,
  author       = {{GitHub}},
  title        = {{GitHub MCP Server}, Release v1.11.0},
  howpublished = {\url{https://github.com/github/github-mcp-server}},
  year         = {2026},
  note         = {README and tool definitions at tag v1.11.0, released 25 August 2026}
}

@misc{github2026scopes,
  author       = {{GitHub}},
  title        = {Scopes for {OAuth} Apps},
  howpublished = {GitHub Docs, \url{https://docs.github.com/en/apps/oauth-apps/building-oauth-apps/scopes-for-oauth-apps}},
  year         = {2026},
  note         = {Accessed October 2026}
}

@misc{github2026pat,
  author       = {{GitHub}},
  title        = {Managing Your Personal Access Tokens},
  howpublished = {GitHub Docs, \url{https://docs.github.com/en/authentication/keeping-your-account-and-data-secure/managing-your-personal-access-tokens}},
  year         = {2026},
  note         = {Accessed October 2026}
}

@article{park2004ucon,
  author  = {Jaehong Park and Ravi Sandhu},
  title   = {The {UCON\textsubscript{ABC}} Usage Control Model},
  journal = {ACM Transactions on Information and System Security},
  volume  = {7},
  number  = {1},
  pages   = {128--174},
  year    = {2004},
  doi     = {10.1145/984334.984339}
}

@inproceedings{birgisson2014macaroons,
  author    = {Arnar Birgisson and Joe Gibbs Politz and {\'U}lfar Erlingsson and Ankur Taly and Michael Vrable and Mark Lentczner},
  title     = {Macaroons: Cookies with Contextual Caveats for Decentralized Authorization in the Cloud},
  booktitle = {Proceedings of the Network and Distributed System Security Symposium (NDSS)},
  year      = {2014}
}

@misc{palumbo2026formal,
  author       = {Nils Palumbo and Sarthak Choudhary and Jihye Choi and Guy Amir and Prasad Chalasani and Somesh Jha},
  title        = {Formal Policy Enforcement for Real-World Agentic Systems},
  year         = {2026},
  eprint       = {2602.16708},
  archiveprefix= {arXiv},
  primaryclass = {cs.CR},
  note         = {Version 3, 8 May 2026}
}

@misc{meng2026cellmate,
  author       = {Luoxi Meng and Henry Feng and Ilia Shumailov and Earlence Fernandes},
  title        = {{ceLLMate}: Sandboxing Browser {AI} Agents},
  year         = {2026},
  eprint       = {2512.12594},
  archiveprefix= {arXiv},
  primaryclass = {cs.CR},
  note         = {Version 2, 4 March 2026}
}

@techreport{jason2004horizontal,
  author      = {{JASON}},
  title       = {Horizontal Integration: Broader Access Models for Realizing Information Dominance},
  institution = {The MITRE Corporation, JASON Program Office},
  number      = {JSR-04-132},
  address     = {McLean, VA, USA},
  year        = {2004},
  month       = dec,
  note        = {Approved for public release}
}

@inproceedings{cheng2007fuzzymls,
  author    = {Pau-Chen Cheng and Pankaj Rohatgi and Claudia Keser and Paul A. Karger and Grant M. Wagner and Angela Schuett Reninger},
  title     = {Fuzzy Multi-Level Security: An Experiment on Quantified Risk-Adaptive Access Control},
  booktitle = {Proceedings of the 2007 IEEE Symposium on Security and Privacy (SP '07)},
  pages     = {222--230},
  year      = {2007},
  doi       = {10.1109/SP.2007.21}
}

@inproceedings{molloy2008trading,
  author    = {Ian Molloy and Pau-Chen Cheng and Pankaj Rohatgi},
  title     = {Trading in Risk: Using Markets to Improve Access Control},
  booktitle = {Proceedings of the 2008 New Security Paradigms Workshop (NSPW '08)},
  pages     = {107--125},
  publisher = {ACM},
  year      = {2008},
  doi       = {10.1145/1595676.1595694}
}

@article{liu2010budget,
  author  = {Debin Liu and L. Jean Camp and XiaoFeng Wang and Lusha Wang},
  title   = {Using Budget-Based Access Control to Manage Operational Risks Caused by Insiders},
  journal = {Journal of Wireless Mobile Networks, Ubiquitous Computing, and Dependable Applications},
  volume  = {1},
  number  = {1},
  pages   = {29--45},
  year    = {2010},
  doi     = {10.22667/JOWUA.2010.06.31.029}
}

@inproceedings{liu2011incentive,
  author    = {Debin Liu and Ninghui Li and XiaoFeng Wang and L. Jean Camp},
  title     = {Beyond Risk-Based Access Control: Towards Incentive-Based Access Control},
  booktitle = {Financial Cryptography and Data Security (FC 2011)},
  series    = {Lecture Notes in Computer Science},
  volume    = {7035},
  pages     = {102--112},
  publisher = {Springer},
  year      = {2012},
  doi       = {10.1007/978-3-642-27576-0_9}
}

@article{salim2013brbac,
  author  = {Farzad Salim and Jason Reid and Uwe Dulleck and Ed Dawson},
  title   = {Budget-Aware Role Based Access Control},
  journal = {Computers \& Security},
  volume  = {35},
  pages   = {37--50},
  year    = {2013},
  doi     = {10.1016/j.cose.2012.11.002}
}

@phdthesis{salim2012thesis,
  author = {Farzad Salim},
  title  = {Approaches to Access Control Under Uncertainty},
  school = {Queensland University of Technology},
  year   = {2012}
}

@article{celikel2009risk,
  author  = {Ebru Celikel and Murat Kantarcioglu and Bhavani Thuraisingham and Elisa Bertino},
  title   = {A Risk Management Approach to {RBAC}},
  journal = {Risk and Decision Analysis},
  volume  = {1},
  number  = {1},
  pages   = {21--33},
  year    = {2009},
  doi     = {10.3233/RDA-2008-0002}
}

@misc{mohammadi2026irreversibility,
  author        = {Bardia Mohammadi and Laurent Bindschaedler},
  title         = {The Irreversibility Budget: Fleet-Level Risk Accounting and Admission Control for Agent Operating Systems},
  year          = {2026},
  eprint        = {2609.00275},
  archiveprefix = {arXiv}
}

@misc{iyer2026racg,
  author        = {Laxmipriya Ganesh Iyer and Rahul Suresh Babu},
  title         = {Capability Minimization as a Safety Primitive: Risk-Aware Causal Gating for Least-Privilege {LLM} Agents},
  year          = {2026},
  eprint        = {2606.13884},
  archiveprefix = {arXiv}
}

@misc{liu2025budgetaware,
  author        = {Tengxiao Liu and Zifeng Wang and Jin Miao and others},
  title         = {Budget-Aware Tool Use Enables Effective Agent Scaling},
  year          = {2025},
  eprint        = {2511.17006},
  archiveprefix = {arXiv}
}

@misc{apono2026guard,
  author       = {{Apono}},
  title        = {Agent Privilege Guard},
  howpublished = {Product page, \url{https://www.apono.io/agent-privilege-guard/}},
  year         = {2026},
  note         = {Accessed 5 October 2026}
}

@inproceedings{dimmock2004trustrisk,
  author    = {Nathan Dimmock and Andr{\'a}s Belokosztolszki and David M. Eyers and Jean Bacon and Ken Moody},
  title     = {Using Trust and Risk in Role-Based Access Control Policies},
  booktitle = {Proceedings of the Ninth ACM Symposium on Access Control Models and Technologies (SACMAT '04)},
  pages     = {156--162},
  publisher = {ACM},
  year      = {2004},
  doi       = {10.1145/990036.990062}
}

@inproceedings{baracaldo2012trustrisk,
  author    = {Nathalie Baracaldo and James B. D. Joshi},
  title     = {A Trust-and-Risk Aware {RBAC} Framework: Tackling Insider Threat},
  booktitle = {Proceedings of the 17th ACM Symposium on Access Control Models and Technologies (SACMAT '12)},
  pages     = {167--176},
  publisher = {ACM},
  year      = {2012},
  doi       = {10.1145/2295136.2295168}
}

@inproceedings{chen2012riskrbac,
  author    = {Liang Chen and Jason Crampton},
  title     = {Risk-Aware Role-Based Access Control},
  booktitle = {Security and Trust Management (STM 2011)},
  series    = {Lecture Notes in Computer Science},
  volume    = {7170},
  pages     = {140--156},
  publisher = {Springer},
  year      = {2012},
  doi       = {10.1007/978-3-642-29963-6_11}
}

@inproceedings{bijon2012risksessions,
  author    = {Khalid Zaman Bijon and Ram Krishnan and Ravi Sandhu},
  title     = {Risk-Aware {RBAC} Sessions},
  booktitle = {Information Systems Security (ICISS 2012)},
  series    = {Lecture Notes in Computer Science},
  volume    = {7671},
  pages     = {59--74},
  publisher = {Springer},
  year      = {2012},
  doi       = {10.1007/978-3-642-35130-3_5}
}

@article{yemini2000marketnet,
  author  = {Yechiam Yemini and Apostolos Dailianas and Danilo Florissi and Gur Huberman},
  title   = {{MarketNet}: Protecting Access to Information Systems Through Financial Market Controls},
  journal = {Decision Support Systems},
  volume  = {28},
  number  = {1--2},
  pages   = {205--216},
  year    = {2000},
  doi     = {10.1016/S0167-9236(99)00069-X}
}

@misc{github2026roles,
  author       = {{GitHub}},
  title        = {Repository Roles for an Organization},
  howpublished = {GitHub Docs, \url{https://docs.github.com/en/organizations/managing-user-access-to-your-organizations-repositories/managing-repository-roles/repository-roles-for-an-organization}},
  year         = {2026},
  note         = {Accessed 9 October 2026}
}

@misc{github2026apptokens,
  author       = {{GitHub}},
  title        = {Generating an Installation Access Token for a {GitHub} App},
  howpublished = {GitHub Docs, \url{https://docs.github.com/en/apps/creating-github-apps/authenticating-with-a-github-app/generating-an-installation-access-token-for-a-github-app}},
  year         = {2026},
  note         = {Accessed 9 October 2026}
}

@misc{aws2026accesslevels,
  author       = {{Amazon Web Services}},
  title        = {Access Levels in Policy Summaries},
  howpublished = {AWS Identity and Access Management User Guide, \url{https://docs.aws.amazon.com/IAM/latest/UserGuide/access_policies_understand-policy-summary-access-level-summaries.html}},
  year         = {2026},
  note         = {Accessed 9 October 2026}
}

@inproceedings{gadelha2018esbmc,
  author    = {Mikhail R. Gadelha and Felipe R. Monteiro and Jeremy Morse and Lucas C. Cordeiro and Bernd Fischer and Denis A. Nicole},
  title     = {{ESBMC} 5.0: An Industrial-Strength {C} Model Checker},
  booktitle = {Proceedings of the 33rd ACM/IEEE International Conference on Automated Software Engineering (ASE)},
  pages     = {888--891},
  publisher = {ACM},
  year      = {2018},
  doi       = {10.1145/3238147.3240481}
}

@misc{farias2024esbmcpython,
  author       = {Bruno Farias and Rafael Menezes and Eddie B. de Lima Filho and Youcheng Sun and Lucas C. Cordeiro},
  title        = {{ESBMC-Python}: A Bounded Model Checker for {Python} Programs},
  year         = {2024},
  eprint       = {2407.03472},
  archiveprefix= {arXiv},
  primaryclass = {cs.SE},
  note         = {Tool demonstration at ISSTA/ECOOP 2024}
}

@misc{tihanyi2023selfhealing,
  author       = {Norbert Tihanyi and Ridhi Jain and Yiannis Charalambous and Mohamed Amine Ferrag and Youcheng Sun and Lucas C. Cordeiro},
  title        = {A New Era in Software Security: Towards Self-Healing Software via Large Language Models and Formal Verification},
  year         = {2023},
  eprint       = {2305.14752},
  archiveprefix= {arXiv},
  primaryclass = {cs.SE},
  note         = {Presented at the 6th ACM/IEEE International Conference on Automation of Software Test (AST 2025)}
}

\end{document}